# Using Schema to Inform Method Design Practices

**Shruthi Sai Chivukula**
Pratt Institute
New York, NY, USA
schivuku@pratt.edu

**Colin M. Gray**
Indiana University
Bloomington, IN, USA
comgray@iu.edu

**ABSTRACT**
There are many different forms of design knowledge that guide and shape a designer's ability to act and realize potential realities. Methods and schemas are examples of design knowledge commonly used by design researchers and designers alike. In this pictorial, we explore, engage, and describe the role of schemas as tools that can support design researchers in formulating methods to support design action, with our framing of method design specifically focused on ethical design complexity. We present four ways for method designers to engage with schema: 1) Systems to operationalize complex design constructs such as ethical design complexity through an A.E.I.O.YOU schema; 2) Classifiers to map existing methods and identify the possibility for new methods through descriptive semantic differentials; 3) Tools that enable the creation of methods that relate to one or more elements of the schema through creative departures from research to design; and 4) Interactive channels to playfully engage potential and new opportunities through schema interactivity.

**CSS Concepts**
• Human-centered computing~Interaction design process and methods

## INTRODUCTION

A variety of forms of design knowledge can be used by practitioners and researchers to inform, shape, and evaluate design action [10,12]. While design knowledge is known to play a critical role in design processes, little is known about how design scholars use forms of knowledge to create design methods [3]. In this pictorial, we explore the intersection of method creation and the generation of guiding structures that inform design action, in the form of *schema*. Schema are "cognitive models, or mental models, that humans create for themselves to help make sense of complex real-world experiences" [12].

Our contribution in this pictorial is providing a visually-oriented account of how schema can be created, iterated upon, and used to inform the creation of new design methods. This process of method creation supported by schema also illustrates how scholars can intentionally activate ethical concerns in the creation of new design methods.

## ETHICAL COMPLEXITY AND METHODS

Methodologies, methods, frameworks, and toolkits have been proposed by HCI and STS scholars that offer support to designers in value inscription, ethical prescription, and building of ethical outcomes (e.g., [3,13]). However, recent scholarship has shown that these methods may lack resonance with the realities of design practice, including a variety of ecological factors that mediate design action [4,5,11].

We describe design methods as "tool[s] that allow designers to support thinking, reflecting and acting upon design activities" [6], while also leveraging an emerging definition of ethics-focused methods, where the "function of the method revealed through this *embedded knowledge* allows designers to convert ethics-focused discovery into design outcomes" [3].

In this pictorial, taking a research through design approach, we present our process of creating schema that allowed us to consider a range of ethical considerations as experienced by practitioners. We describe how we leveraged these schema to create new methods, including ones that map the designer's ecology, leverage the use of ethical dilemmas, and evaluate existing methods. The primary contribution of the pictorial is not the methods themselves, but rather how we iteratively and reflexively created schema that helped us explore and operationalize the space of ethical complexity in the design of methods.

## SETTING THE STAGE

**Design Schema**
We build upon Nelson and Stolterman's [12] definition of a design schema as a "means for representing holistic concepts, ideas, and fundamental knowledge in visual form." Unlike scientific schema, we position the creation and use of schemas as a designerly practice, whereby schema can support the method designer in better understanding the design space, provide a vocabulary to structure or explain problem space traversal, and indicate gaps, opportunities, or combinations of approaches that can be discovered through schema use. We focus on the utility of this approach for design scholars creating new methods to increase ethical engagement, but similar processes or concepts could be taken on by practitioners as well as they seek to appropriate existing methods or build new ones to address concerns in their practice.

**Different Types of Schemas We Built**
We constructed two different types of schemas. First, we used *semantic differentials* as a structure to describe the descriptive qualities of a method. And second, we used *structural schema* that resemble models to connect the creation of the ethics-focused methods (our particular scope in the context of our larger research aims) with the broader literature around ethical complexity in HCI practice.

> Schema are "knowledge structures, cognitive structures, and or strategies" [12] that allow us to represent, discuss, expand, and imagine design opportunities and processes.

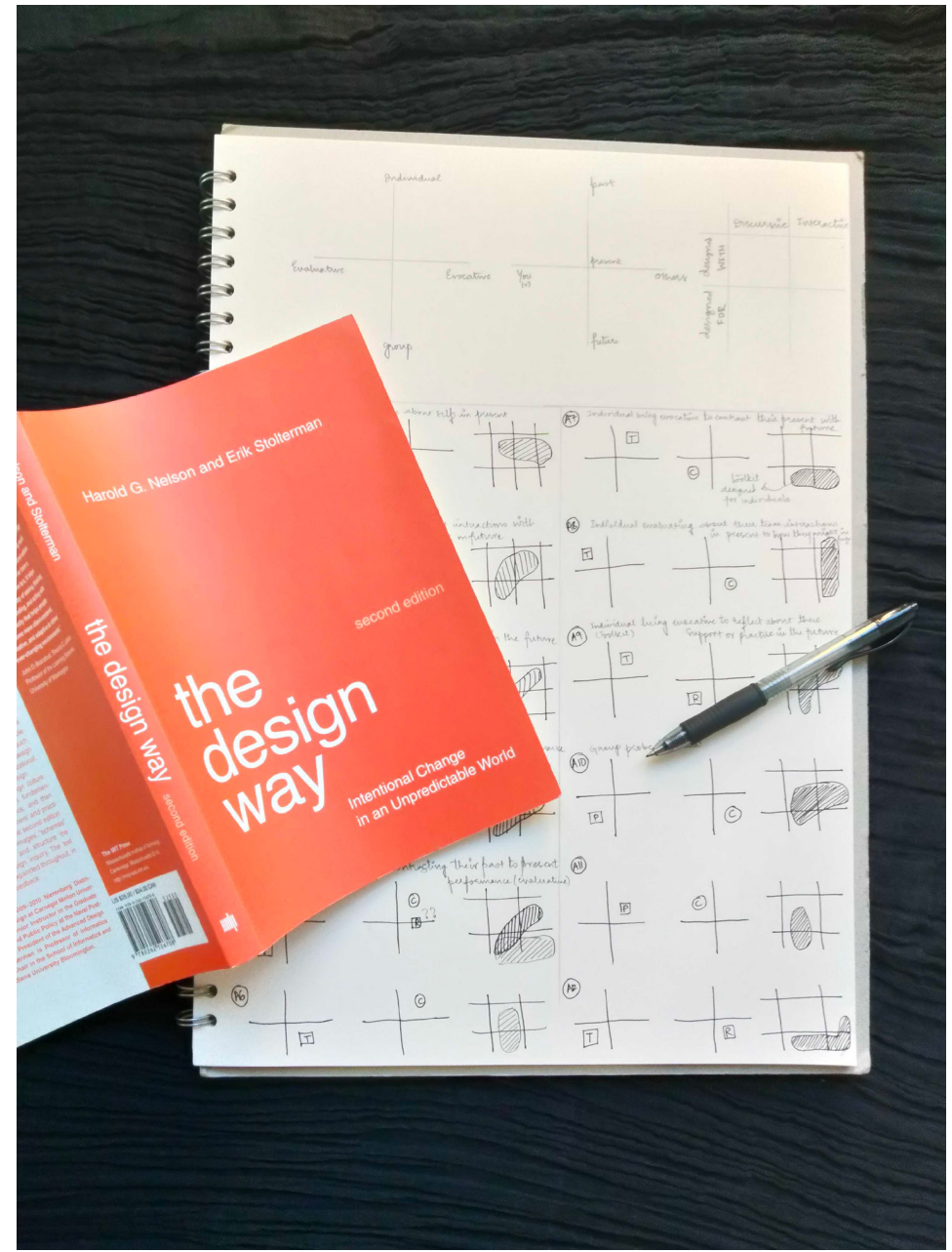

## PROJECT OVERVIEW

Our method design process spread across a timeline of six months, beginning with a divergent creation of concepts for methods that could reveal ethical complexity in technology practice. At the end of each set of divergent activities, we generated a pool of ideas and created visual structures—or schema—to move forward with our design process.

These schema helped us to formulate a vocabulary to describe potential methods and their possible variations. We first created a range of methods and formulated an initial structure through the A.E.I.O.YOU schema. We then iterated on variations of methods, creating classifier or descriptive schema that distilled specific method constraints for further consideration.

We then created digital mock-ups and final prototypes of our proposed methods. The schema and method creation process is detailed in the following indicated pages, contributing to our overall process.

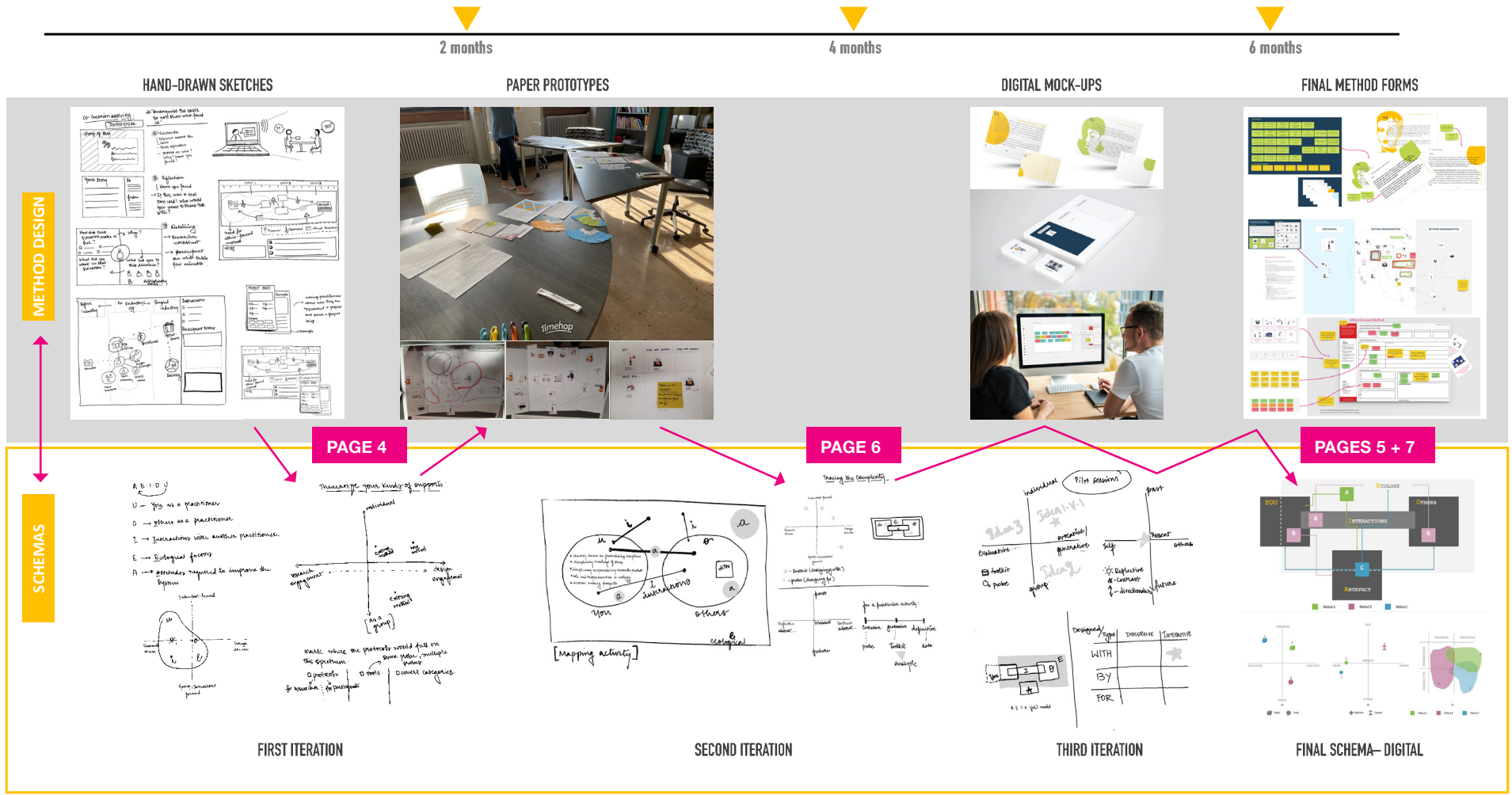

2 months | 4 months | 6 months

HAND-DRAWN SKETCHES | PAPER PROTOTYPES | DIGITAL MOCK-UPS | FINAL METHOD FORMS

METHOD DESIGN

PAGE 4 | PAGE 6 | PAGES 5 + 7

SCHEMAS

FIRST ITERATION | SECOND ITERATION | THIRD ITERATION | FINAL SCHEMA— DIGITAL

# IDEAS → SCHEMA

**BUILDING A.E.I.O.YOU**

We began by sketching a range of potential methods to address and describe ethics in a practitioner's everyday work. An affinity mapping of all the methods we created resulted in three main themes: 1) the individual practitioner and their interactions with other practitioners; 2) ecological factors that practitioners are embedded within; and 3) attitudes or artifacts that they may need to support more ethical work practices. This affinity mapping gave us an overarching structure and resulted in a system-oriented schema. This became our first structured schema—which we call **A.E.I.O.YOU**—that helped us to package and describe a frame for many potential new methods that could address ethical concerns.

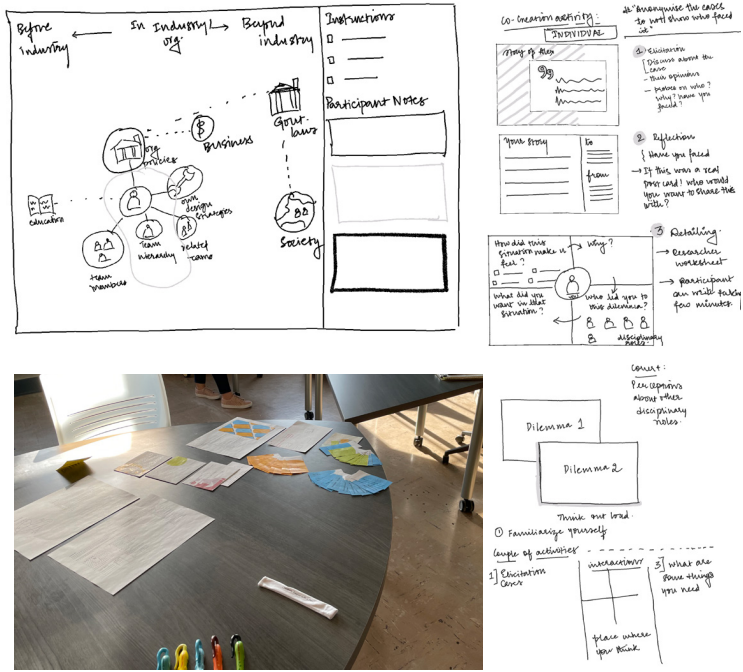

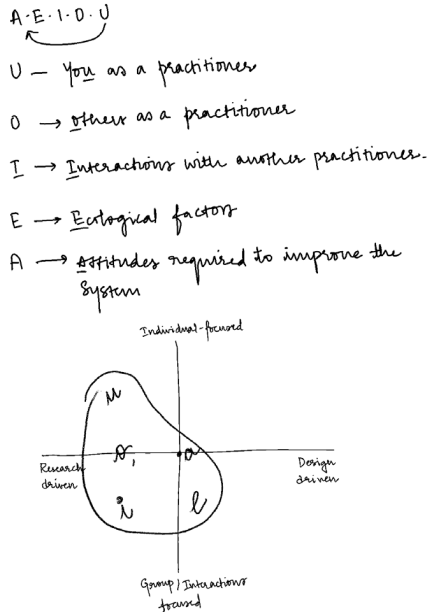

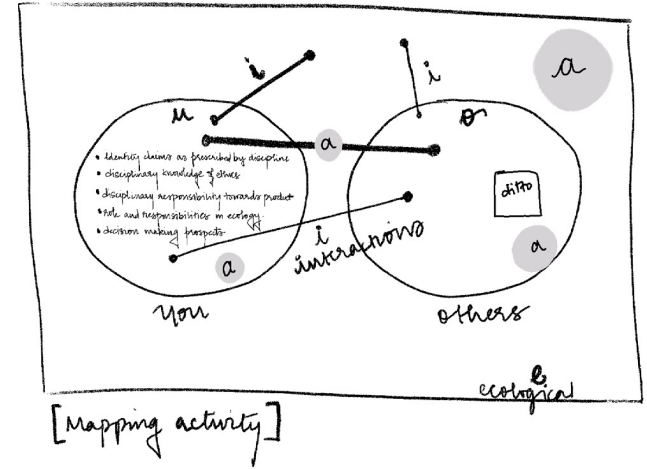

**SKETCHES**

*Initial sketches and paper prototypes of a range of methods to address and describe ethics in a practitioner's everyday work.*

**ITERATION #1**

*First iteration of A.E.I.O.YOU created from affinity of sketches, overlapped with a first interaction of semantic differential schema.*

**ITERATION #2**

*Second iteration of A.E.I.O.YOU independent of other schemas to create a stand-alone version.*

## A.E.I.O.YOU SCHEMA

**A**    <u>Artifacts</u> for support for practitioner's ethical engagement.

**E**    <u>Ecological Factors</u> and complexity that the practitioner is a part of.

**I**    <u>Interactions</u> with other practitioners during ethical engagement.

**O**    <u>Other practitioners</u> and their responsibility in ethical decision making.

**YOU**    <u>YoU</u> refers to individual practitioners, their ethical awareness, responsibility, and action, within and beyond ecological boundaries.

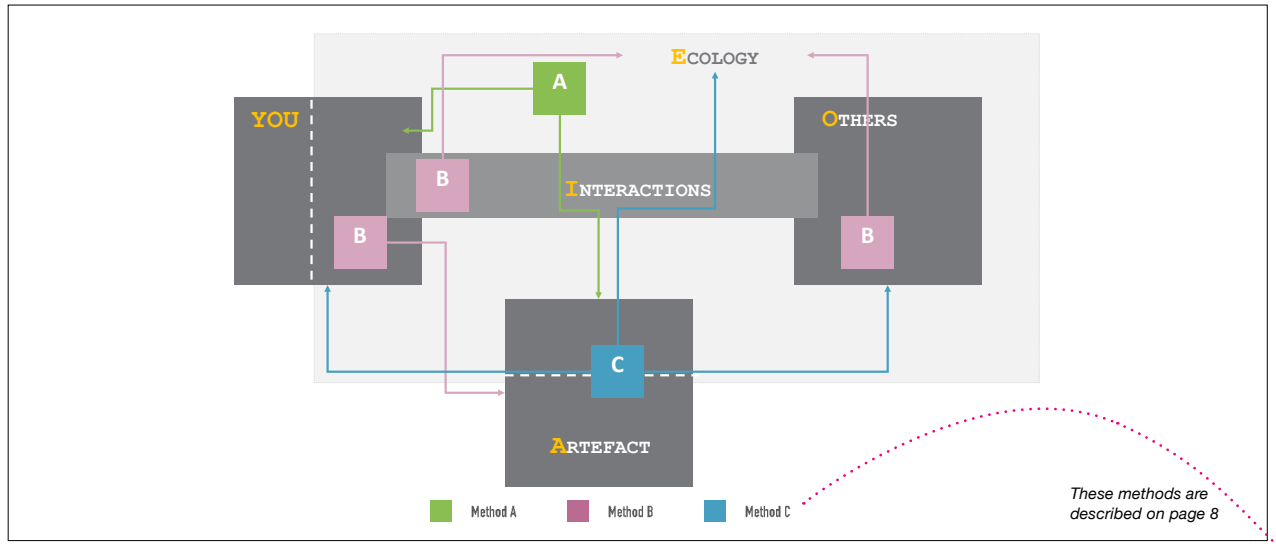

*These methods are described on page 8*

---

**Connections to literature**
- Existing ethics-focused methods [3,14]

<u>ARTIFACTS include</u>
- **Methods to support ethics-focused design work**
- Frameworks or practices to follow ethical responsibility.
- Attitudes to improve ethical action.

**A**  *Bold text indicates a focus for our research project*

---

**Connecting to Literature**
- Describing ethical design complexity [5]
- Foregrounding soft resistance [15]

<u>ECOLOGICAL FACTORS include</u>
- Industry standards, processes, and policies.
- Business setup and goals.
- **External factors influencing ethical decision making.**

**E**

---

**Connecting to Literature**
- Using Values Levers and related organizational practices [13]

<u>INTERACTIONS include</u>
- **Conversations or partnerships in team(s) and with stakeholder(s).**
- Collaborations with external clients.
- Coordination with internal and external teams.

**I**

---

**Connecting to Literature**
- Differences in disciplinary roles [1]

<u>OTHERS includes</u>
- **Team members**
- **Other professional roles**
- Business Stakeholders or Clients
- Users

**O**

---

**Connecting to Literature**
- Constructing identity claims [2]
- Noticing, reflecting, and reacting to ethics [9]

<u>YOU includes</u>
- **Practitioners**
- Educators
- Students

**YOU**

**ITERATING ON CLASSIFIER SCHEMA**

We created the schema through multiple iterations, including reshaping the schema focus and language and the medium (from sketches on a digital tablet or paper interaction to digitizing for final production).

**First Iteration**

We first described methods based on the possibility for individual or group-based engagement (A) and if they were design-focused or research-focused (B), seeking to identify methods which engaged designers in creating artifacts versus methods that were discursive in nature.

**Second Iteration**

We then built on this schema, repeating the first schema [(C) = (A) + (B)] and creating:

- the A.E.I.O.YOU schema (D)
- introducing axes of temporality (E) and reflection/contrast (F)
- creating a scale to mark if a particular activity is evocative, generative, or definitive (G) in various forms such as a probe, toolkit, or data (H)

**Third Iteration**

We then iterated further, creating:

- creating (I) from a combination of (A) + (G) + (H)
- introducing (J), (K) [based on (F)], and (E) to create a new axis between involved personnel and temporality
- introducing (L) to create a matrix across involvement and mode of engagement [connected to tangible forms of (G)], using (D) to identify elements from the A.E.I.O.YOU schema

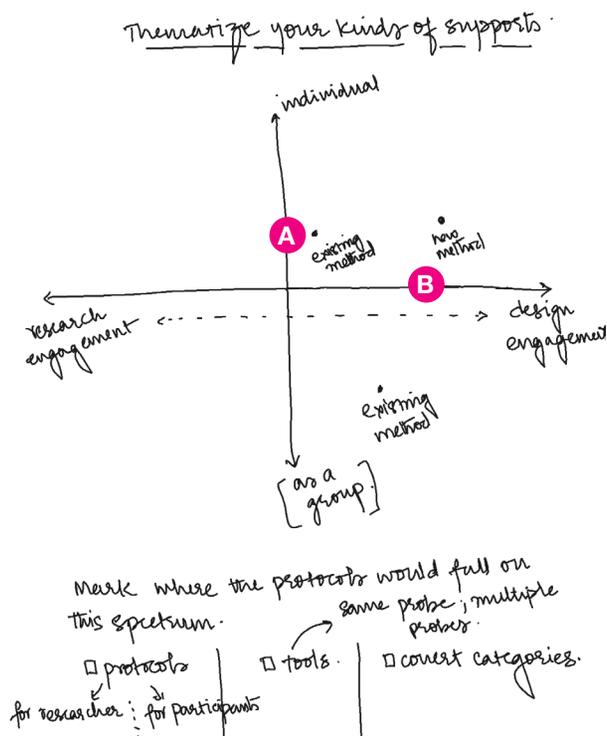

**ITERATION #1**

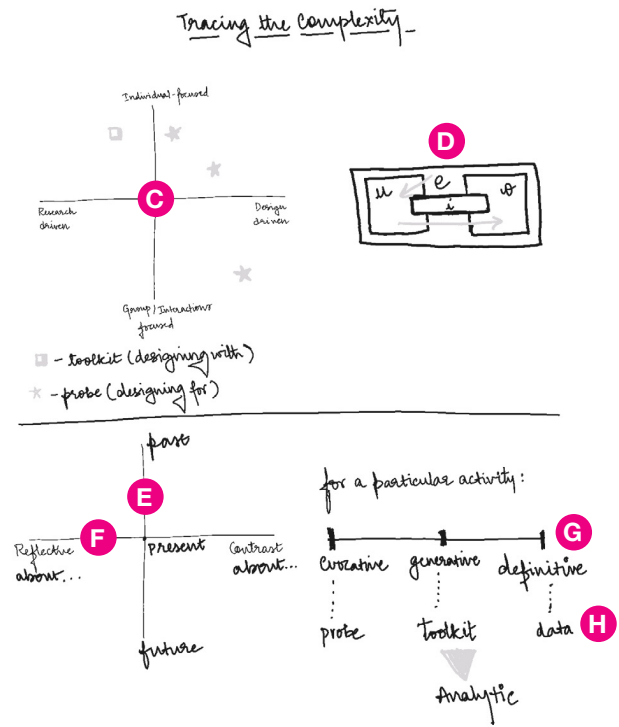

**ITERATION #2**

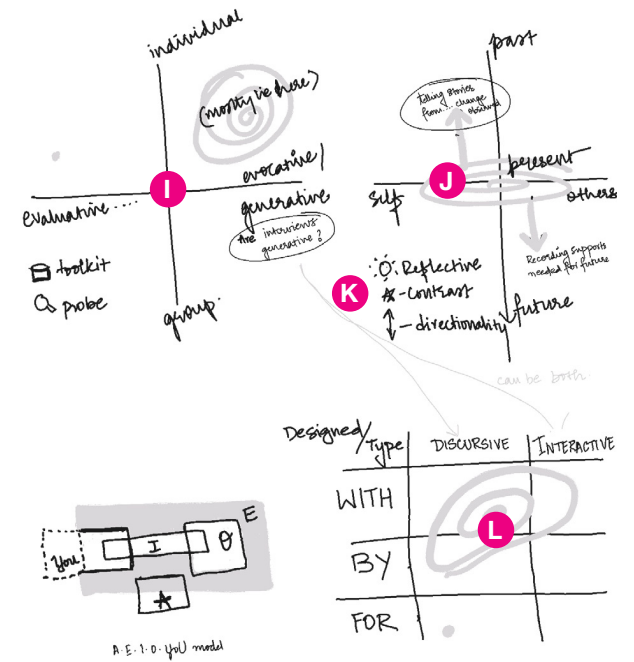

**ITERATION #3**

**FINAL CLASSIFIER SCHEMA**

### Schema A: Size x Function

This schema describes how a method could engage different sizes of groups across different types of functions. The function axis indicates whether a method's purpose is primarily *evaluative*, where a method is designed with a primary goal of assessing an artifact, situation, practitioners' mindset, shared scenarios, or knowledge, or *evocative*, where a method is designed with the primary goal of assisting practitioners in articulating tacit knowledge about ethical considerations, ethical supports, or a practitioners' mindset.

### Schema B: Involved Personnel x Temporality

This schema describes how a method could engage practitioners in their own or other practitioner's situations, scenarios, ecology, knowledge, roles, discipline, ethical valence, or ethical support. The temporality axis describes the time frame of knowledge used in the method and the personnel axis reflects or contrasts the engagement of a practitioner's own ethical awareness, engagement, and action or the experiences of others across the provided time frames.

### Schema C: Involvement x Mode of Engagement

This schema describes how methods relate to practitioner involvement and the mode of engagement. "Designed with" practitioners activities included practitioner's engagement with the provided toolkits or probes to create artifacts or narrate stories through two modes of engagement: *Discursive*, where the practitioner is engaged in a conversational act using the designed probes, and/or *Interactive*, where the practitioners interact with the method by creating artifacts using the designed toolkits.

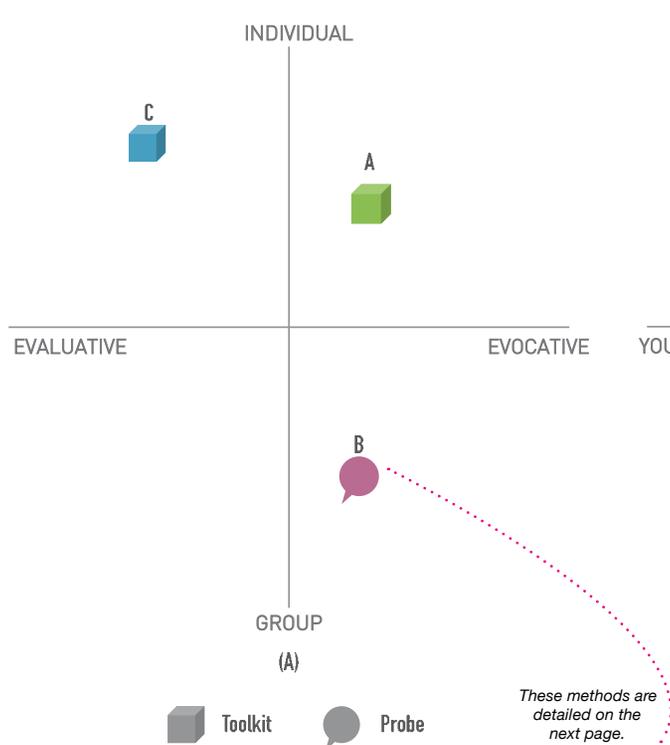

(A)

*These methods are detailed on the next page.*

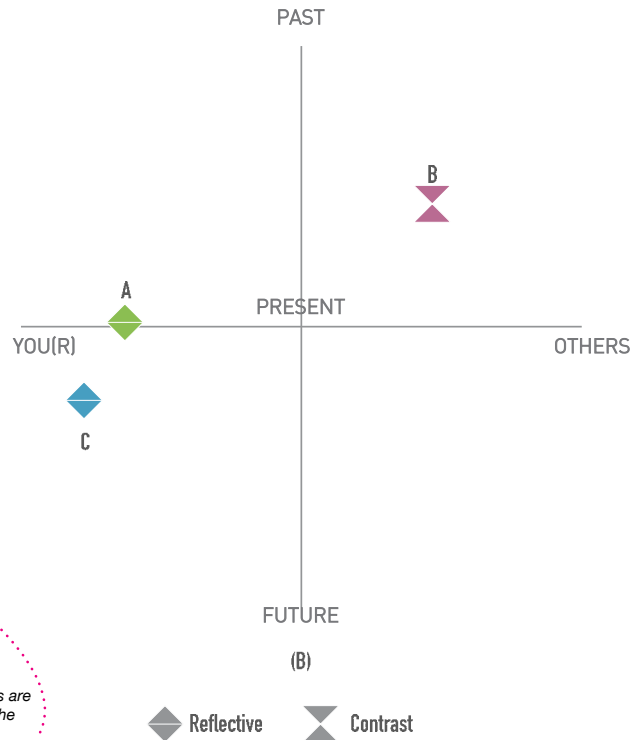

(B)

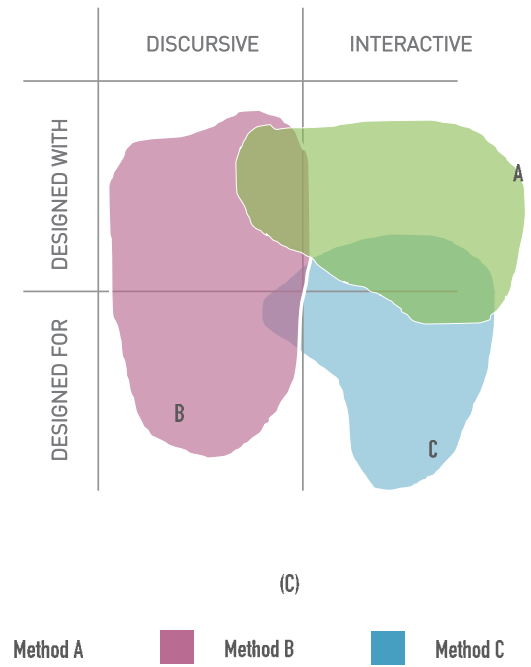

(C)

**METHODS WE DESIGNED USING THE SCHEMA**  *These methods are mapped to the schema on the previous page*

### A. "Tracing the Complexity"
This method encourages practitioners to map their experiences of ethical complexity, including individual, organization, and societal (beyond organizational) dimensions that impact the practitioner's interactions with other practitioners and their decision making towards the designed technological product.

| | |
|---|---|
| **Schema A** | Individual + Evocative + Toolkit |
| **Schema B** | Reflective + Present + Your Own |
| **Schema C** | Interactive + Designed for Practitioners |

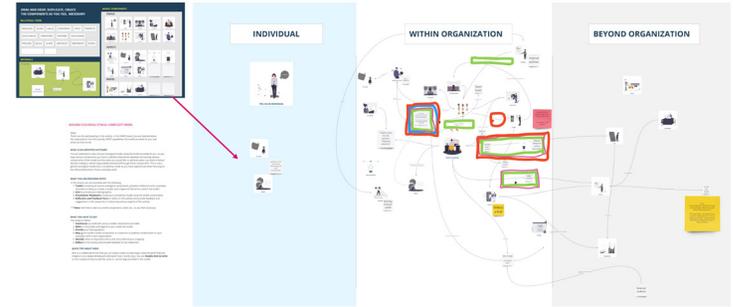

---

### B. "Dilemma Postcards"
This method engages practitioners to share their stories based on a list of ethical dilemmas provided to them or reflect and react to ethical dilemmas stories shared by other practitioners, as they face these ethical dilemmas in their decision making in their everyday work.

| | |
|---|---|
| **Schema A** | Individual + Evocative + Probe |
| **Schema B** | Reflective + Past/Present + Your Own |
| **Schema C** | Discursive + Designed for practitioners |

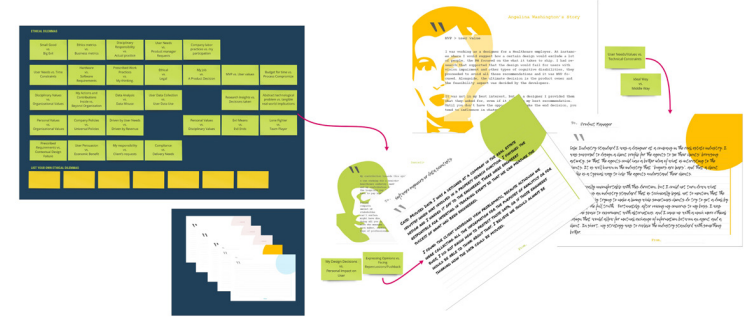

---

### C. "Method Heuristics"
This method introduces practitioners to an existing ethics-focused method to evaluate it for its performance, i.e. how does the method enable ethical decision making in practitioners' everyday work, and prescription, i.e. how does the method guide through decision making.

| | |
|---|---|
| **Schema A** | Individual + Evaluative + Toolkit |
| **Schema B** | Reflective + Present/Future + Your Own |
| **Schema C** | Interactive + Designed for and with Practitioners |

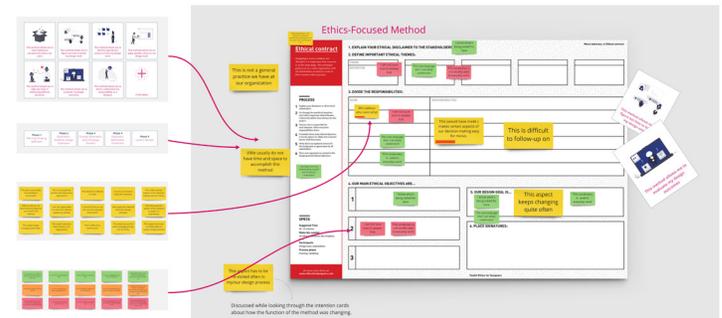

*METHOD*  *CONNECTIONS TO SCHEMA*  *FINAL METHOD FORM*

## INTERACTIVELY PLAYING WITH SCHEMA

*schema frame interactive play*

*descriptive/classifier schema*

*physical manipulatives*
- **P** Probe
- **T** Toolkit
- **R** Reflective
- **C** Contrast

*possible interactions with physical manipulatives*

*sketches from interacting with the schema*

*possibilities of method creation*

SCHEMA → IDEAS

**SCHEMA-INFORMED DESIGN PRACTICES**
Throughout our research process, schema played four primary roles: 1) aiding us in scoping and shaping the research space; 2) as heuristics to build a vocabulary to understand what methods could be built; 3) as a tool to support structured ideation; and 4) as an early evaluation tool.

**Scoping and Shaping the Research Space**
Schema provided us with a visual tool to shape and identify the part of the entire landscape of defining ethical complexity we could address given the project timeframe and other constraints. The projection of all the variations of methods on the final descriptive schema resulted in a visual tool to identify potential future method combinations. Conversely, the schema also provided us with constraints such as not being able to engage a group of individuals in using methods due to project timeline and other logistics.

**Heuristics**
Schema provided us visual and verbal guidelines and designerly language to describe potential methods. The three methods we created (A, B, and C) enabled us to tackle different perspectives based on the A.E.I.O.YOU schema, and the variations under each of these broad intentions were strengthened by using the schema as a guidebook to describe and differentiate among different methods or variations.

**Ideation Tools**
Schema were visual tools to map, filter, and combine different axes that informed the creation of a design frame, supporting ideation and sketching of new variants. As we updated and iterated on the descriptive schemas, ideation became a playful activity where we "mixed and matched" to consider potential ways to engage practitioners through new methods. Schema gave us new ideas about different forms (probes or toolkits), functions (evocative or evaluative), intentions (reflect or contrast), and other facets of potential methods. Apart from designing new variants of the methods, the visual nature of the schema helped us to consider different combinations and sequences of the methods we created.

**Evaluation Tools**
Schema helped us to continuously evaluate and reflect on the differences among variations of the methods and if a certain designed method fell under our scope for this research project. For example, we designed a variation of Method A that had more generative properties; however, we chose not to focus on generative qualities in this research project. Similarly, we re-evaluated variants we had already designed, asking if they were more focused towards a "generative" focus, which allowed us to eliminate such variants for our study.

**CONCLUSION**
In this pictorial, we have presented two types of schema: 1) *A.E.I.O.YOU*, a structured topic-based visual structure designed to represent a landscape of research in ethical complexity in HCI practice; and 2) *Descriptive Schema*, a semantic differential form iteratively created to describe a range of possible methods. We describe the process of generating these schemas and their potential utility in identifying opportunities for creating new methods. Throughout the pictorial, we illustrate how these schemas have supported us, as design researchers, to organize our ideation process, identify divergent opportunities, guide the creation of a range of methods, and playfully interact with intangible aspects of method creation.

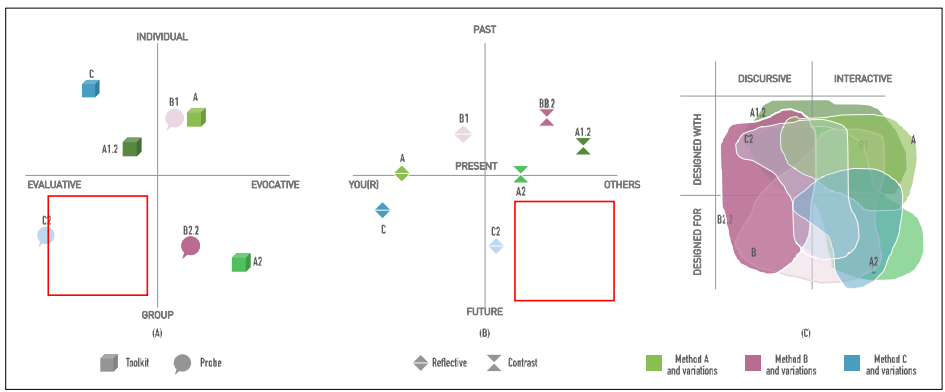

*using schema as a heuristic to scope the research space and provide a designerly language*

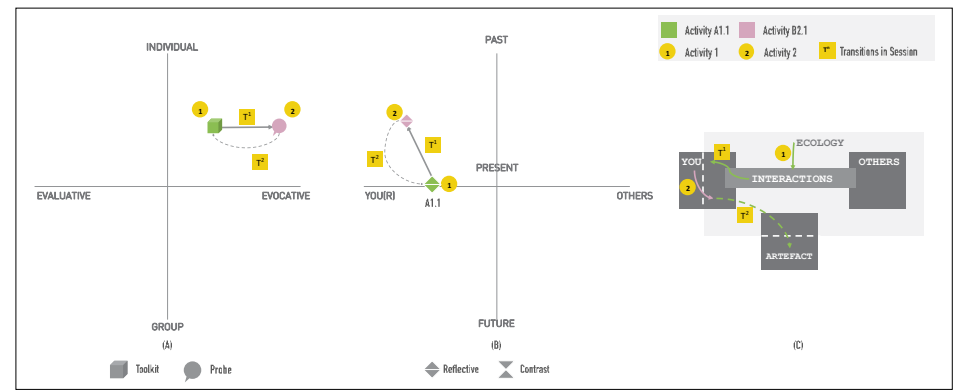

*using schema as an ideation tool to imagine different combinations of methods and interactive sequencing opportunities*

**ACKNOWLEDGMENTS**
Left blank for anonymous review.